# Distributed Detection in Sensor Networks with Limited Range Sensors [*]

Erhan Baki Ermis, Venkatesh Saligrama

Department of Electrical and Computer Engineering

Boston University, Boston, MA 02215

Email: {ermis, srv}@bu.edu

November 4, 2018

## Abstract

We consider a multi-object detection problem over a sensor network (SNET) with limited range sensors. This problem complements the widely considered decentralized detection problem where all sensors observe the same object. While the necessity for global collaboration is clear in the decentralized detection problem, the benefits of collaboration with limited range sensors is unclear and has not been widely explored. In this paper we develop a distributed detection approach based on recent development of the false discovery rate (FDR). We first extend the FDR procedure and develop a transformation that exploits complete or partial knowledge of either the observed distributions at each sensor or the ensemble (mixture) distribution across all sensors. We then show that this transformation applies to multi-dimensional observations, thus extending FDR to multi-dimensional settings. We also extend FDR theory to cases where distributions under both null and positive hypotheses are uncertain. We then propose a robust distributed algorithm to perform detection. We further demonstrate scalability to large SNETs by showing that the upper bound on the communication complexity scales linearly with the number of sensors that are in the vicinity of objects and is independent of the total number of sensors. Finally, we deal with situations where the sensing model may be uncertain and establish robustness of our techniques to such uncertainties.

## 1 Introduction

The design and deployment of sensor networks (SNET) for distributed decision making pose fundamental challenges due to energy constraints and environmental uncertainties. While power and energy constraints limit collaboration

---

[*]This research was supported by ONR Young Investigator Award N00014-02-100362, Presidential Early Career Award (PECASE), and NSF Award CCF-0430993

among sensors nodes, some form of collaboration is necessary to overcome uncertainty and meet reliability requirements of the decision making process. The general question of dealing with distributed data in the context of detection has been an active topic of research (see [6, 17, 21–24] and references therein).

In this paper we focus on the problem of distributed detection of localized events, sources or abnormalities, and seek to devise a distributed detection strategy that satisfies false alarm and communication cost constraints. The problem of localized detection arises naturally in many setups, e.g. whenever there are multiple objects in a surveillance area and the sensing range of each sensor is significantly small relative to the surveillance area. For example, a number of objects generate a spatially confined signal field and the sensors sample the field at their locations as illustrated in Figure 1 (b). Preliminary work along these lines have been presented in some of our earlier papers [8, 9, 25]. We note here that this work is the first step toward identifying the set of sensors that are proximal to a given object. Another interesting subject of research is how to fuse the information from proximal sensors to determine precise object location. This latter objective can possibly be accomplished through decentralized or distributed fusion techniques. However, we do not pursue this objective here.

The problem under consideration complements others wherein noisy information about a single event is measured by the entire network (global information problems). For global information problems, researchers have investigated several architectures ranging from fusion centric to ad-hoc consensus based approaches [2, 3, 6, 12, 14, 16, 17, 21–24]. Although we do not discuss this problem in our paper, many distributed inferencing problems can be sub-divided in to two problems: sensor selection, to select sensors in the vicinity of a target, followed by decentralized/distributed processing of information among the selected sensors. Our paper is related to the former problem of sensor selection and is described in [9].

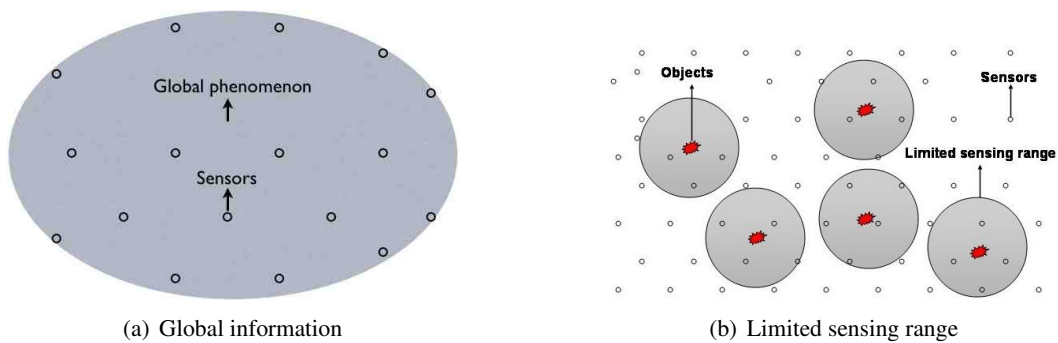

(a) Global information  (b) Limited sensing range

Figure 1: Decentralized Detection vs. Localized Detection: In decentralized detection the sensors observe a single global phenomenon, whereas in localized detection the sensors observe multiple local phenomena.

The problem of distributed detection of localized phenomena is closely related to the multiple hypotheses testing problems considered in statistical literature [13]. In multiple hypotheses testing problems a set of observations is given with each observation coming from one of two distributions and the objective is to associate each observation with its correct distribution. This is different from binary hypothesis testing problems with multiple observations, where all the observations come from the same hypothesis. For instance, in Figure 1 (a), the observations of all the sensors are generated by a single hypothesis, i.e. presence or absence of global phenomenon, and the hypotheses set consists of two hypotheses. On the other hand, in Figure 1 (b), the observation of each sensor is generated by its own set of hypotheses, and the hypotheses set for the whole network can be as large as $2^m$ hypotheses, where $m$ is the number of sensors.

Although false alarm probability is commonly controlled as the reliability criterion in classical hypothesis testing problems, in multiple hypotheses testing problems it invariably results in poor detection performance [1, 4, 5]. In order to compensate for poor detection performance, probability of false alarm can be controlled in a test-wise manner, a method known as uncorrected testing. The uncorrected testing can be thought of as optimizing a Bayes risk criterion for some object density (sparsity level). Here the risk can be the number of errors. Whenever the actual object density differs from the implicitly assumed density, the error rates degrade significantly. Therefore we consider a recently introduced reliability criterion: the BH procedure for controlling false discovery rate (FDR) [4, 11, 19]. Briefly, FDR is the expected ratio of number of false positives to total number of declared positives. This relaxation and the associated BH procedure has been shown to adapt to unknown levels of sparsity [1]. The best rule to reduce the errors is the Bayes risk optimal policy that is tuned to the correct object density and the BH procedure tracks the performance of Bayes Oracle risk policy under assumptions of monotonicity of the distribution under significant hypothesis. As seen from Figure 2, the BH procedure tracks the performance of Bayes oracle risk policy utilizing either the object density or the distribution under positive hypothesis.

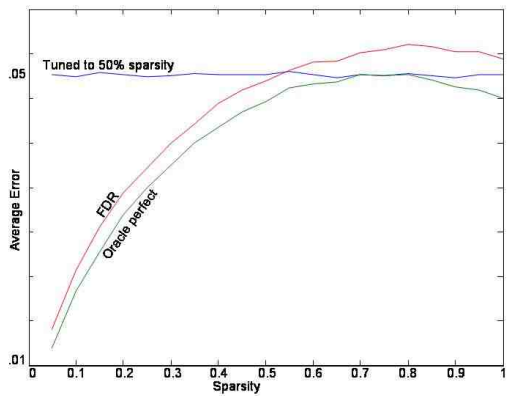

Figure 2: FDR adapts to unknown sparsity levels: 100 samples with distributions $N(0, 1)$ under $H_0$ and $N(0, 3)$ under $H_1$. Average errors were found using Monte-Carlo simulations.

Nevertheless, the BH procedure suffers from many drawbacks in the context of SNETs. First, the distribution under significant hypothesis does not satisfy monotonicity conditions. Moreover interpreting such single-dimensional conditions in multi-dimensional settings is unclear. On the other hand, in most SNET scenarios the observed dis-tributions are at least partially known and the BH procedure does not exploit this knowledge. We first develop a transformation that exploits this knowledge to satisfy these monotonicity conditions. This transformation while controlling the FDR at the same level as the BH procedure dramatically improves detection performance. The transformation is then shown to apply to multi-dimensional observations, thus providing a natural extension to the existing single dimensional procedure. This is particularly useful in situations where the sensed information includes object features in a multi-dimensional space. Next we present a distributed algorithm for SNETs whose communication cost, in terms of the broadcast messages, scales with the number of significant sensors in the SNET, and not the total number of sensors. A very interesting implication of this work is that *corresponding to an FDR threshold the communication cost grows in proportion to the actual number of events, sources or abnormalities while achieving the same centralized performance.* In many situations we have: (a) partial knowledge of the distributions under significant hypothesis; (b) estimates for the mixture distribution of the sensor observations; (c) computational errors introduced particularly in multi-dimensional settings. To address these situations we develop a robust extension to our procedure. Robustness in an important attribute in SNETs. The object intensity distribution generally follows a power law. Therefore, the signal measured at the sensor is the superposition of signals from all the unknown objects.

The organization of the paper is as follows: in Section 2 we present an overview of the setup. We then describe ideal and non-ideal sensing models, describe the false discovery rate, and present the general formulation of the problem. In Section 3 we describe the BH procedure, which controls the false discovery rate, and explain its suboptimal nature. We then develop the domain transformed BH (DTBH) scheme and show that it outperforms BH procedure in terms of detection power given the same FDR constraint. We further present the solution to the problem with ideal sensing model via a distributed DTBH algorithm and present the scalability property. In Section 4 we perform the robustness analysis of the DTBH algorithm. We show that under certain conditions DTBH procedure controls false discovery rate to within a factor of $\epsilon$ when the ideal sensing model is relaxed. In Section 5 we present simulations and discuss some interesting results.

## 2  Setup

We consider a non-Bayesian setting where an unknown number of objects are distributed on a sensor field of $m$ sensors. We assume no prior information on the number of objects, and their potential locations. Objects are observed by a SNET in which the sensor nodes are distributed uniformly. We wish to identify, via distributed strategies, the set of sensors that have an object in their sensing range.

To simplify details pertaining to communication complexity we assume a broadcast model whereby a message from a sensor is broadcast to the entire network. The communication complexity is the aggregate number of mes-sages broadcast by the sensors in the SNET until algorithm termination.

We consider an object centric scheme in which the objects generate a signal field over the sensor network and the sensors sample the field at their locations. In this scheme the positive hypothesis ($H_1$) for a sensor is the event that the sensor is within the effective region of an object, and the null hypothesis ($H_0$) is the event that the sensor is outside the effective region of all objects.

The observation vector is denoted by $Y = (Y_s : s \in \mathcal{S})$, where throughout the paper $\mathcal{S}$ represents the set of sensors that form the SNET, and $Y_s$ represents the collection of measurements taken by sensor $s \in \mathcal{S}$. The realization of observation vector $Y$ is denoted by $y = (y_s : s \in \mathcal{S})$. For definiteness we focus on the case when $Y$ has a continuous distribution. The cummulative probability distribution (resp. density) function of the observation vector $Y_s$ at sensor $s$ under each hypothesis $H^s = H_i$, $i = 0, 1$ is denoted by $G_{is}(\cdot)$ (resp. $g_{is}(\cdot)$), where $H^s$ denotes the hypothesis at sensor $s$. Note that both the CDF and PDF can be suitably described for multidimensional observations. We assume a general structure on the problem in the sense that $G_{1s}$ belongs to a class of distributions $\mathcal{G}_1$, and $G_{0s}$ belongs to a class of distributions $\mathcal{G}_0$. With a slight abuse of notation, we will use these families for distributions and densities where it will not cause confusion. Let $\mathcal{S}_0 = \{s \in \mathcal{S} : H^s = H_0\}$ with cardinality $m_0$ and $\mathcal{S}_1 = \{s \in \mathcal{S} : H^s = H_1\}$ with cardinality $m_1$. Here both $m_0$ and $m_1$ are unknown and the object locations are assumed to be arbitrary, i.e. not necessarily uniformly distributed.

## 2.1 Mathematical Modeling

We describe mathematical models for ideal and non-ideal sensing. The ideal sensing model accounts for situations where objects can be sensed only if a sensor is within a fixed range of object.

**Ideal Sensing Model:** When we mention ideal sensing model we mean that each object has a fixed range in which it generates a uniform signal and outside this range the object has no signal. Each sensor samples the field at its location, of course with some measurement noise:

$$
\begin{aligned}
H^s &= H_0: \ y_s \sim g_{0s} \\
H^s &= H_1: \ y_s \sim g_{1s}
\end{aligned}
$$

For example in a linear model, one may consider

$$
\begin{aligned}
H^s &= H_0: \ y_s = n_s \\
H^s &= H_1: \ y_s = \theta_t + \nu_s
\end{aligned}
$$

where $n_s$ and $\nu_s$ are noise variables with known distributions, and $\theta_t$ is the uniform signal of object $t$ in the vicinity of which sensor $s$ is present. Note that the class of distributions $\mathcal{G}_0$ and $\mathcal{G}_1$ are singletons in the ideal case. In this case the following factorization holds:

$$\Pr\{y_s \mid \{H^s\}_{s\in\mathcal{S}}\} = \Pr\{y_s \mid H^s\}$$

**Non-Ideal Sensing Model:** Here we assume that the observed signal from each object decays as a function of distance between the object and the observing sensor. Therefore the object signal is no longer constant within the region around the object, and is no longer zero outside this region. As a consequence, sensors outside the effective region also observe a signal from each object. Here the received signal carries uncertainty due to unknown fading gains as well as observation noise, thus we assume knowledge of the models to within families that are not signletons:

$$H^s = H_0 : y_s \sim \tilde{g}_{0s} \in \mathcal{G}_0$$
$$H^s = H_1 : y_s \sim \tilde{g}_{1s} \in \mathcal{G}_1$$

An instantiation for this case is

$$H^s = H_0 : y_s = \sum_{t':d(s,t')>d_0} \frac{\theta_{t'}}{(d(s,t')+1)^\alpha} + n_s$$
$$H^s = H_1 : y_s = \frac{\theta_t}{(d(s,t)+1)^\alpha} + \sum_{t':d(s,t')>d_0} \frac{\theta_{t'}}{(d(s,t')+1)^\alpha} + \nu_s$$

where $d(s,t)$ denotes the distance between sensors $s$ and an object $t$, $\alpha$ denotes the decay exponent, and $d_0 > 0$ and the constant one in the denominator has been added to eliminate singularities. In this model the observations are correlated under both hypotheses, and the factorization presented for the ideal sensing model does not hold. Figure 3 illustrates these models.

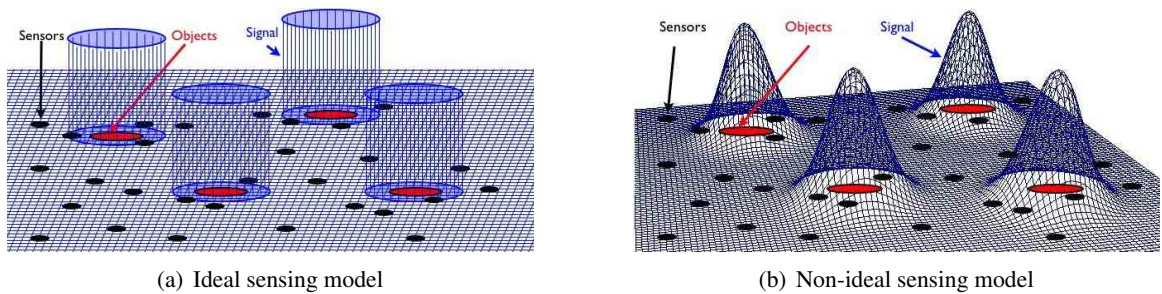

(a) Ideal sensing model  (b) Non-ideal sensing model

Figure 3: Ideal and non-ideal sensing models.

To deal with the non-ideal sensing model, we relax the assumption that $\mathcal{G}_0$ and $\mathcal{G}_1$ are singletons. We deal with this case from a robustness perspective, i.e., as a perturbation of the ideal sensing model, in the upcoming parts of the paper.

## 2.2 Formulation and Objective

Before we proceed further we present the following table, which describes important variables in our discussion. Here $m$ is the number of samples (or sensor nodes) known in advance. The observable random variable $R$ is the total number of sensors that decide positive hypothesis, and the unobservable random variable $V$ is the total number of sensors falsely decide positive hypothesis. The false alarm and miss probabilities are associated with the random

|           | Declared $H_0$ | Declared $H_1$ | Total     |
|-----------|----------------|----------------|-----------|
| True $H_0$ | $U$            | $V$            | $m_0$     |
| True $H_1$ | $T$            | $S$            | $m - m_0$ |
| Total     | $m - R$        | $R$            | $m$       |

variables $V$ and $T$ respectively in the above table. As we discussed in the introduction, the solutions based on false alarm control cannot adapt to various levels of sparsity. In the appendix, we show via information-theoretic arguments (by appealing to Fano lower bound) that asymptotically the worst-case error probability can be bounded from below by the conditional entropy, obtained by substituting a uniform prior on objects and their locations. Consequently, either the miss rate or the false alarm rate is bounded from below by half the conditional entropy. The corresponding theorem is stated below:

**Theorem 2.1** *Suppose, $u(Y^m)$ is any strategy that maps the sensor observations to object locations. Then,*

$$\gamma_w = \min_u \max_{H^s} \left( Pr\{V \geq 1 \mid \{H^s : s \in \mathcal{S}\}\} + Pr\{T \geq 1 \mid \{H^s : s \in \mathcal{S}\}\} \right) \geq \gamma_b$$

*where*

$$\gamma_b \doteq \min_u Pr(V \geq 1) + Pr(T \geq 1) \geq \Phi(H^s \mid Y_s) - \frac{1}{N}$$

*where $\Phi(\cdot \mid \cdot)$ is the conditional entropy and $\gamma_b$ is the Bayes uniform risk obtained by assuming a uniform distribution of objects at $m$ locations. It follows that there exists no decision strategy for which both false alarm and miss probability can simultaneously be smaller than $\Phi(H^s \mid Y_s)/2$.*

**Remark:** $H^s$ is a binary random variable and so its entropy (or conditional entropy) is always smaller than one. Nevertheless, depending on measurement noise at each sensor $\Phi(H^s \mid Y_s)$ could be arbitrarily close to one.

In the FDR [4] formulation we control the worst case expected ratio of $V/R$, i.e.

$$FDR = \max_{\{H^s\}_{s \in \mathcal{S}}} E\{V/R \mid \{H^s\}_{s \in \mathcal{S}}\}$$

For simplicity of notation, from here on we will write $FDR = E\{V/R\}$ and $\Pr\{\cdot \mid \{H^s\}_{s \in \mathcal{S}}\} = \Pr\{\cdot\}$ whenever it is clear from the context. For completeness, we show that $FDR$ can be expanded as follows:

$$\begin{aligned} FDR &= E\{V/(V+S)\} = E\{V/R\} \\ &= E\{V/R|R>0\}\Pr\{R>0\} + E\{V/R|R=0\}\Pr\{R=0\} \\ &= E\{V/R|R>0\}\Pr\{R>0\} \end{aligned}$$

The last equality follows from the convention that $E\{V/R|R=0\} = 0$. This is intuitively pleasing, because if no sensors declare presence of object within their sensing radius then no false discoveries are committed. In this work we seek to use the FDR framework to perform detection in the sensor network problem that has been laid our earlier. We wish to devise a distributed detection method that controls the FDR at desired levels. The general form of our problem is to minimize the expected miss rate subject to false alarm rate and communication cost constraints.

## 3 FDR Control and Domain Transformation

In this section we will take a close look at the BH procedure, expose its weaknesses, develop a domain transformation to improve on its shortcomings, and present the solution to the detection problem with ideal sensing model.

**BH procedure:** For completeness we first describe the BH procedure which is also illustrated in Figure 4.First, the so called p-values are computed. The $p$ value of an observation $y_s$ is a **non-unique** transformation that generates a uniform distribution under null hypotheses. One such transformation is $p_s = P(y_s) = 1 - G_0(y_s)$ but other transformations that are related to $\alpha$-level significance regions are possible [13, 20]. The p-values are then ordered and the largest index $i_{max}$, such that $p_i \leq \frac{i}{m}\gamma$ is chosen. All the indices smaller than $i_{\max}$ are labeled significant.

Let $Y_{0s} \sim G_{0s}$ (resp. $Y_{1s} \sim G_{1s}$) be the observed random variable under null (resp. positive) hypothesis at sensor $s$. Define $P_{0s} = P(Y_{0s})$, and similarly $P_{1s} = P(Y_{1s})$ and let $F_{0s}$ and $F_{1s}$ be their corresponding distribution functions, i.e. $P_{0s} \sim F_{0s}$ and $P_{1s} \sim F_{1s}$. The family $\mathcal{G}_0$ is transformed to a new family $\mathcal{F}_0$ and $\mathcal{G}_1$ is transformed to a new family $\mathcal{F}_1$ by this transformation. The following theorem and its proof are presented in [4]. We state the theorem without proof and refer the reader to [5] for further details.

**Theorem 3.1** *For independent test statistics under null hypothesis, and for any configuration of positive hypotheses,*

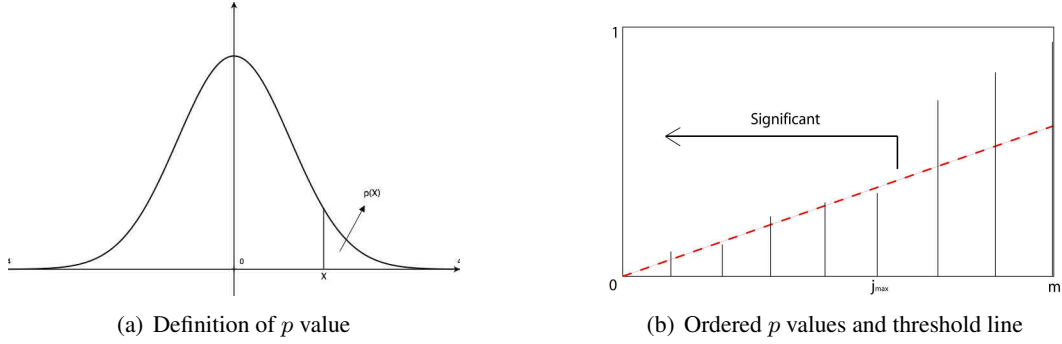

(a) Definition of $p$ value  (b) Ordered $p$ values and threshold line

Figure 4: BH procedure

*the BH procedure controls the FDR at level $\gamma m_0/m$, where $m_0$ is the number of true null hypothesis and $m$ is the number of observations.*

The main idea behind the proof of Theorem 3.1 lies in the simple fact that the observations under null hypotheses are independent, and that the $P_{0s} \sim U(0,1) \ \forall s \in \mathcal{S}$.

**Shortcomings of BH Procedure:** First, the BH procedure performs well only when the realizations of $P_1$ are monotonic and clustered near zero, which requires a certain structure on the distribution of observations. This issue is particularly problematic for multi-dimensional distributions since it is unclear how to reduce the multi-dimensional observations to 1-Dimension. Moreover, simple strategies such as projections to 1-D do not result in monotonicity and clustering around zero. Second, the BH procedure does not take into account the knowledge of the probability distributions that generate the samples under $H_1$. The focus primarily is to reduce false positives and there is no control over the miss rate. Indeed by suitable transformations it maybe possible to realize the clustering around zero. Third, the BH procedure does not lend itself easily to decentralized implementation. More specifically, the BH procedure is a last crossing procedure wherein the largest $p$ value smaller than its corresponding threshold must be found. This requires searching among $\gamma m$ $p$-values, which scales with the number of sensors in the network. In this section we address both these issues by using exact knowledge of distributions. We devise a first crossing procedure that achieves the detection power of last crossing procedure without the communication overhead that scales with the number of sensors. In Section 4 we extend this approach to cases where the distributions are known partially.

An example where clustering around zero is not guaranteed follows:

**Example:** Consider two Gaussian random variables with $Y_{0s} \sim N(0,1)$ and $Y_{1s} \sim N(0,.01)$ for $s = 1 \ldots m$, and consider the FDR constraint $\gamma = .05$. Assume that we are given $m$ $p$ values calculated via the transformation $p(y_s) = 1 - G_0(y_s)$. The goal is to select the samples of $P_{1s}$ from a mixture of samples subject to FDR constraint, $\gamma$. In this example most of the realizations of $P_{1s}$ are close to 0.5 rather than 0. Note, however, that the BH procedure seeks for $P_{1s}$ samples that are less than or equal to .05. Therefore it will not declare any sample of $P_{1s}$ as significant, and BH procedure results in zero detections. To overcome this problem, consider the following transformation on the random variables $P_{0s}$ and $P_{1s}$:

$$\hat{P_{is}} = |1 - 2P_{is}|, i = 0, 1, s = 1 \ldots m \tag{1}$$

Since $P_{0s}$ is uniformly distributed in (0,1), a little algebra shows that $\hat{P_{0s}}$ is also uniformly distributed in (0,1). Therefore we know that if we use BH procedure on the new set of $p$ values we can still control the FDR at .05. Observe, however, that most of the realizations of $\hat{P_{1s}}$ are now close to 0. When the BH procedure is performed on this new set of $p$ values, more of the observations coming from $H_1$ will be declared as significant, thus the detection power is increased. ■

Such cases where realizations of $P_1$ are away from zero can arise in many situations. Another example is when $Y_{0s}$ and $Y_{1s}$ are exponential random variables with parameters $\lambda_{0s} = 2$ and $\lambda_{1s} = 1$ respectively. In this case, using the $p$ value definition $p(y_s) = 1 - G_0(y_s)$, the realizations of $P_{1s}$ are close to 1. A similar strategy to the previous example can be employed to resolve the issue again. In fact, a different definition of $p$ value can be used to evade this problem all together. However, in more general cases, for example when the null distribution is a mixture distribution, finding a suitable $p$ value definition may not be evident or simple.

More interesting examples arise when we consider *multi variate distributions*. We give an example of this nature in the sequel. *Generally speaking, computationally convenient definitions of $p$ values do not generate $P_0$ and $P_1$ distributions suitable for BH procedure's direct application.* Therefore we must make use of the knowledge of probability distributions that generate samples under positive hypotheses and transform the $p$ values accordingly.

## 3.1 Domain Transformed BH Procedure

We now develop a method to overcome the issues we observed on the BH procedure. The main idea of this section is motivated by our example, and is based on the following insight: recall that all that is necessary for BH procedure to control FDR is (a) the observations be independent under null hypotheses, (b) $p$ values be distributed $U(0, 1)$ under null hypotheses. Assume that we can find a transformation $\mathcal{T}$ such that $\mathcal{T}(P_1)$ is concentrated near 0, and $\mathcal{T}(P_0) \sim U(0, 1)$. Then, (1) since $\mathcal{T}(P_0) \sim U(0, 1)$ we are not interfering with conditions (a) and (b) above, hence BH procedure will control the FDR with the new $p$ values. (2) $\mathcal{T}$ maps samples of $P_1$ to near 0, and generates a new data set that is more suitable for BH procedure. We restate (1) as a proposition below and formally examine (2) in the upcoming sections.

**Proposition 3.2** *Let $p_1, p_2, \ldots, p_m$ be a set of p values such that $P_0 \sim U(0, 1)$ and $\mathcal{T} : (0, 1) \to (0, 1)$ be a function. If $\mathcal{T}(P_0) \sim U(0, 1)$, then BH procedure controls FDR at desired levels when applied to $\mathcal{T}(p_1), \mathcal{T}(p_2) \ldots \mathcal{T}(p_m)$, and we say that $\mathcal{T}$ is measure invariant with respect to $P_0$.*

Proof: Since the distribution under null hypothesis is preserved to be $U(0,1)$, following the proof of theorem 3.1 gives the result stated in the proposition. ∎

The proposed transformation is a reorientation of the $p$ domain. It depends on the distribution of observations under positive hypotheses, but not the realizations themselves. Its main features are: (a) preserves uniform distribution of $p$ values under null hypothesis, which implies that the FDR constraint is not violated; (b) maps an arbitrary PDF of $p$ values to a monotonically decreasing one, which leads to improved detection rates (see Figure 5). We will see how to extend this idea to the multi-dimensional setting in the following section.

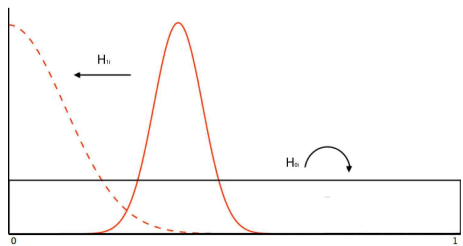

Figure 5: Illustration of domain transform: a monotonically decreasing density is obtained.

## 3.2 Transformation of $p$ Domain and Multi-dimensional FDR control

Suppose, without loss of generality[1], assume that the null distribution is $P_{0s} \sim U^k(0,1)$. Suppose, $f_1(\cdot)$ is the corresponding distribution under $H_1$ supported in the cube.

**Special Case when $f_1$ is nowhere constant:** Let $p(y) = \Pr\{p \mid f_1(p) \geq f_1(y)\}$ be the transformation. This transformation involves computing volume of level sets under $f_1$. The fact that this transformation satisfies the desired objective will be shown as part of the general transformation later in this section. We next describe an example to illustrate the utility of the transformation in a multi-variate problem.

**Example:** Let $P_{0s} \sim U^2(0,1)$, and $P_{1s} \sim F_1(0,1)$ where $F_1(0,1)$ is a 2 dimensional circularly symmetric distribution centered around $(.5,.5)$ supported on $[0,1]^2$ as depicted in Figure 6. Under the proposed domain transformation, $\{x : f_1(x) > f_1(P)\}$ for some $P$ is always a disk, $D$, centered around $(.5,.5)$, the edge of which goes through $P$. The transformed $p$ value is the area of this disk, i.e. $\hat{P} = \int_D 1 dx$, $D = \{x : f_1(x) \geq f_1(P)\}$. This transformation is depicted in Figure 7 (a). Now consider a different transformation which computes the area radially outside the observed value and maps that area to $\hat{P}$, which is depicted in Figure 7 (b).

Note that in the corresponding 1-D space, our proposed transformation separated the distribution of $\hat{P}_{1s}$ from uniform much better than the radial transformation did. Furthermore the resulting $p$ values are concentrated near 0

---
[1]A uniform distribution achieving transformation is as follows: Let $g(y_1, y_2)$ be a 2-D distribution. Define, $p(y_1) = \int_{y_1}^{\infty} \int_{-\infty}^{\infty} g(t,s) dt ds$ and $p(y_2 \mid y_1) = \int_{y_2}^{\infty} g(s \mid y_1) ds$ where $g(s \mid y_1)$ denotes the conditional distribution 2nd dimension given the first.

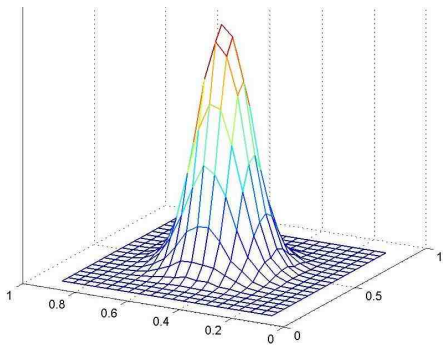

Figure 6: Non-normalized density of 2 dimensional $P_{1s}$

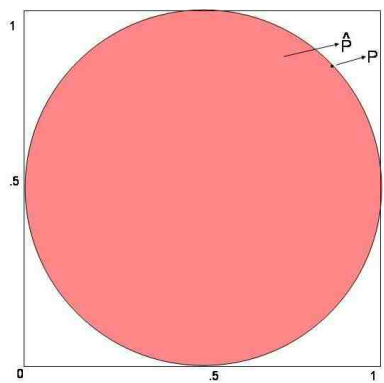
(a) Domain transformation with respect to $F_1(0, 1)$

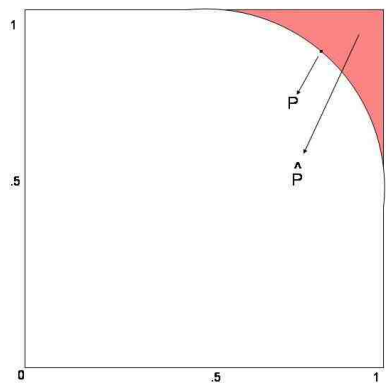
(b) Radial transformation

Figure 7: Illustration of transformation from 2 dimensions to single dimension

for the proposed transformation as depicted in Figure 8. Another method that has been used in multidimensional cases can be found in [7], where FDR is applied separately for each dimension with varying thresholds so that one can still preserve global FDR control. It turns out that this transformation is sub-optimal as well and fails when the distributions are not separable in any single dimension.

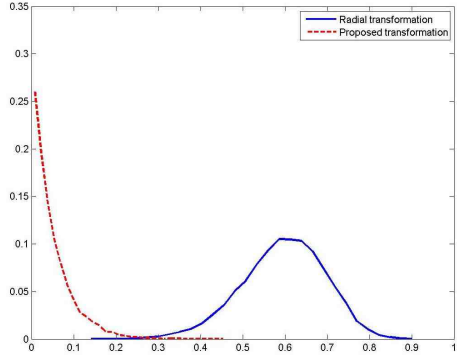

Figure 8: Non-normalized density of 1 dimensional $\hat{P}_{1s}$

To establish our results we consider the general 1-D setting for simplicity of exposition. These definitions and related results apply to multi-dimensional setting without any changes where the definition of $p$ value satisfies $P_{0s} \sim U^k(0,1)$. For simplicity of notation, we will drop the sensor subscript $s$, adopting $P_0$ for $P_{0s}$, $P_1$ for $P_{1s}$, and similarly for densities and distributions. In fact, the transformation $\mathcal{T}$ we define is associated with a sensor and it should be denoted $\mathcal{T}_s$, however we omit the subscript. Now let $f_1(\cdot)$ be the PDF of $P_1$, which exists since $P_1$ is a continuous random variable. Define the transformation, $\mathcal{T}$, as follows:

1. Let $y_{max} = \sup_x \{f_1(x)\}$. Define $\alpha_\mu(y) = E_U[I_{\{f_1(x) \geq y\}}(x)]$ and $\beta_\mu(y) = E_\mu[I_{\{f_1(x) \geq y\}}(x)]$ for $y \in (0, y_{max})$ where $\mu$ is the measure of $P_1$; i.e. $\mu(A) = \int_A f_1$. Intuitively, $\alpha_\mu(y)$ captures the length of the set $\{x : f_1(x) \geq y\}$, and $\beta_\mu(y)$ captures the probability of $P_1$ falling in the set $\{x : f_1(x) \geq y\}$.

2. Generate a new measure: $\hat{\mu}(0, \alpha_\mu(y)) = \beta_\mu(y) \ \forall y \in (0, y_{max})$. If $\alpha_\mu(y)$ has a jump at $y = y_0$ from $a$ to $b$, then set
$$\hat{\mu}(0, z) = \frac{\beta_\mu(y_0^-) - \beta_\mu(y_0^+)}{b - a}(z - a) + \beta_\mu(y_0^+)$$
for $z \in (a, b)$, which corresponds to a conditionally uniform distribution in $(a,b)$. Let $\hat{f}_1(\cdot)$ be the corresponding density of $\hat{\mu}$.

3. Generate the transformed random variable $\hat{P} = \mathcal{T}[P]$ as follows: For $P \in (0, 1)$ find $Y = f_1(P)$; then find the set $S = \{x : \hat{f}_1(x) = Y\}$ and choose $\hat{P}$ randomly from $S$.

Various elements involved in this definition can be understood through Figure 9. Note that the above general definition reduces to a simple expression for nowhere constant $f_1$: $\hat{P} = E_U[I_{\{f_1(x) \geq f_1(P)\}}(x)] = \int_A 1 dx, \ A = \{x :$

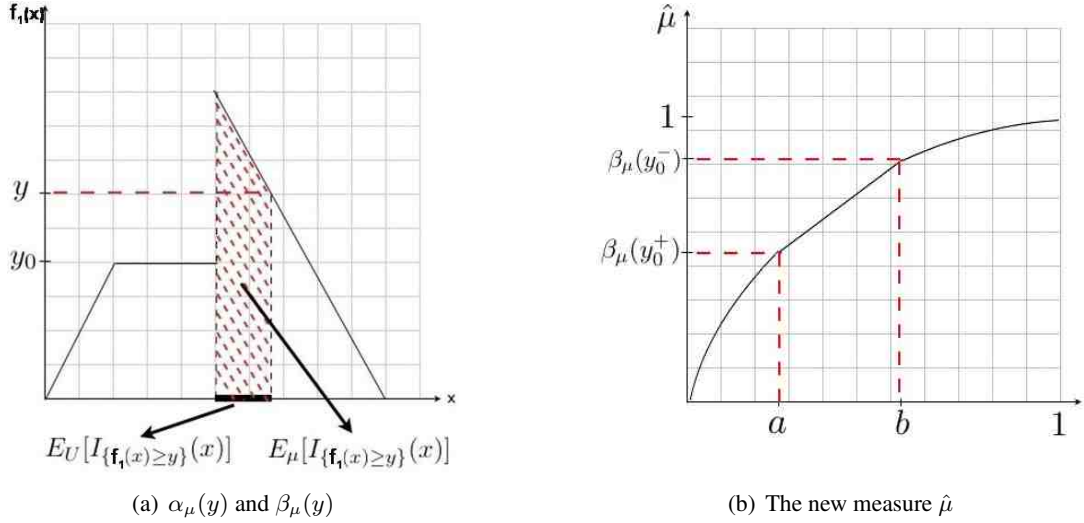

(a) $\alpha_\mu(y)$ and $\beta_\mu(y)$  (b) The new measure $\hat{\mu}$

Figure 9: Illustration of transformation elements

$f_1(x) \geq f_1(P)\}$. To firmly establish the validity of these definitions we need to ensure that $\beta_\mu$ can be regarded as a distribution function and that $\hat{\mu}$ is absolutely continuous with respect to the Lebesgue measure, which we do so in the appendix.

In summary the Domain Transformed BH (DTBH) procedure involves: (a) Apply the transformation $\mathcal{T}$ to realizations of the random variables $P_0$ and $P_1$. (b) Follow the BH procedure

We must now establish two facts pertaining to $\mathcal{T}$. First we must show that $\mathcal{T}$ is measure invariant with respect to the distribution of $P_0$, i.e. $\mathcal{T}[P_0] \sim U(0,1)$. Recall that we need this condition for DTBH procedure to control FDR at desired levels. Next we need to show that realizations of $\mathcal{T}[P_1]$ are indeed clustered near 0. We establish the the former via the following two results:

**Proposition 3.3** *$\mathcal{T}$ is a measure invariant transformation with respect to $U(0,1)$.*

Proof: By definition $\mathcal{T}$ maps countable sets to singletons. Furthermore sets of non-zero Lebesque measure are mapped to sets of same Lebesque measure, hence the uniform distribution is preserved. ∎

**Proposition 3.4** *The DTBH procedure controls FDR at the same level as BH procedure.*

Proof: By Proposition 3.3 $\mathcal{T}[P_0] \sim U(0,1)$. Then the result follows from Proposition 3.2. ∎

Now that we have shown DTBH controls FDR at desired levels, we are left to show that realizations of $\mathcal{T}[P_1]$ are clustered near 0. Formally, we show that $\mathcal{T}[P_1]$ has a monotonically decreasing density. This result will be of great importance in proving improved performance of DTBH procedure over BH procedure.

**Proposition 3.5** *$\mathcal{T}$ converts an arbitrary continuous density of $P_1$ to a monotonically decreasing density over $(0,1)$; i.e. $\hat{f}_1(\hat{p})$ is monotonically decreasing in $\hat{p}$.*

Proof: See appendix.

## 3.3 Performance Comparisons

Here we define threshold strategies, and show that with the domain transformation the optimal detection rule is a threshold strategy. We also show that DTBH procedure has stronger detection power in comparison to BH procedure.

**Definition 3.6** *Assume a partitioning problem of a set of observations $X = \{x_1, x_2, \ldots, x_m\}$ into two subsets, $X_1$ and $X_2$ such that $X_1 \cap X_2 = \phi$ and $X_1 \cup X_2 = X$. A threshold strategy is one that computes a threshold $t(x_1, x_2, \ldots, x_m)$, and partitions $X$ into two sets: $X_1 = \{x \in X : x \leq t(x_1, x_2, \ldots, x_m)\}$ and $X_2 = \{x \in X : x > t(x_1, x_2, \ldots, x_m)\}$.*

We have the following theorem when the object locations are all equally likely on the sensor network but nevertheless the parameter governing the likelihood of an object at a location is unknown.

**Theorem 3.7** *Let all object locations be equally likely on the sensor network, i.e. $Pr\{H^i = H_0\} = Pr\{H^j = H_0\}$, $\forall i, j : 1, \ldots, m$, and let $f_0(\cdot) = U(0, 1)$. If $f_1(\cdot)$ is a monotonically decreasing PDF, a thresholding strategy is optimal.*

Proof: See appendix. ∎

Before proceeding any further, the term *stochastically larger* [18] must be introduced: We say that the random variable $X$ is stochastically larger than the random variable $Y$, denoted $X \geq_{st} Y$, when $F_X(a) \leq F_Y(a)$ for all $a$.

**Lemma 3.8** *Let $X_1..X_n \in (0, 1)$ be $n$ independent random variables with common density function $f_X$ and let $Y_1..Y_n \in (0, 1)$ be $n$ independent random variables with common density function $f_Y$. Also, let $X_{(i)}$ and $Y_{(i)}$ denote the $i^{th}$ smallest of $X_1..X_n$ and $Y_1..Y_n$ respectively. If $F_X(t) \geq F_Y(t) \ \forall t \in (0, 1)$, then $Y_{(i)} \geq_{st} X_{(i)}$.*

Proof: See appendix. ∎

The important implication of this lemma is captured in the following theorem.

**Theorem 3.9** *For any given set of p values with known distributions and any integer k, the probability of declaring the first k p values as significant is larger under the DTBH procedure than the BH procedure.*

Proof: Let $\hat{P}_1 = \mathcal{T}[P_1]$. By construction of the transformation, the density of $\hat{P}_1$, $\hat{f}_1$, dominates the density of $P_1$, $f_1$. In other words, $\hat{P}_1 \leq_{st} P_1$. Therefore, the results of the lemma 3.8 apply to random variables $P_1$ and $\hat{P}_1$.

First, assume that the observations contain only samples from $H_0$. Since the random variable $\mathcal{T}[P_0]$ is stochastically equivalent to $P_0$, the probability of declaring $k$ of them significant is equal for all $k$ with both procedures. Let $p'$ be the $k^{th}$ $p$ value. Next, when the samples of $H_1$ are added one by one, probability of the index of $p'$ increasing to $k+1$ is larger with addition of samples from $\mathcal{T}[P_1]$ in comparison to addition of samples from $P_1$. This is because $P_1$ is stochastically larger than $\mathcal{T}[P_1]$. But, since $p' \leq k\gamma/m$ implies $p' \leq (k+1)\gamma/m$ the DTBH procedure increases the probability of a $p$ value being declared as significant. Furthermore, this argument is valid for all $k \leq m$, since $\mathcal{T}$ converts an arbitrary continuous density of $P_1$ to a monotonically decreasing one, which concludes the proof. ∎

Figure 10 demonstrates the detection power of DTBH procedure in comparison to that of the BH procedure. The former is uniformly stronger than the latter.

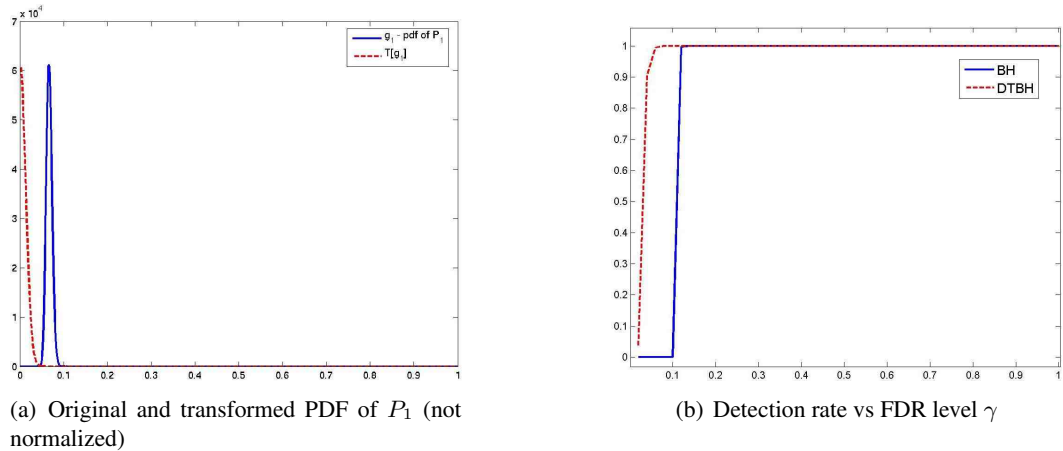

(a) Original and transformed PDF of $P_1$ (not normalized)

(b) Detection rate vs FDR level $\gamma$

Figure 10: Comparison of detection performance for BH and DTBH procedures.

Next we describe an algorithmic solution to the distributed detection problem with ideal sensing model case. The distributed solution is based on the DTBH procedure, and can be seen as a distributed implementation of DTBH with communication constraints.

## 3.4 Distributed Detection with Ideal Sensing Model

The DTBH procedure consists of two main parts. The first part is the domain transformation, which does not require any communication between sensor nodes. This is because the distribution of the random variable $P_{1s}$ is available at sensor $s \in \mathcal{S}$, and the domain transformation depends only on this information. Therefore, it can be applied at each sensor node locally. The second part of the DTBH procedure is the BH procedure itself, which requires ordering of $p$ values. Since ordering of $p$ values is costly in terms of communications, we use a sequential method to accomplish the linearly increasing thresholding of BH procedure. See [15] on how single bit information can be transmitted efficiently to implement this sequential method. As discussed earlier we consider communication complexity by the number of broadcast messages.

**Distributed DTBH Algorithm:** At iteration $t$ each sensor keeps a threshold variable $l(i_t) = i_t\gamma/m$ and a bit counter $count_t$. Initialize $i_1 = 1$ and $count_0 = 0$. Then:

0. Each sensor performs domain transformation

1. Sensor $j$ decides $H_{1j}$ if $p_j \leq l(i_t)$ and $H_{0j}$ otherwise. If decided $H_{1j}$, announces to the network if it has not done so at iterations $1\ldots t-1$

2. Assume $R_t$ sensors decide $H_1$ and declare to the network. Set $i_{t+1} = i_t + 1$ & $count_t = count_{t-1} + R_t$

3. If $count_t \geq i_t$ mark iteration $t_{max}$

4. If $i_t = m$ or $R_t = 0$ label sensors that declare $H_1$ until iteration $t_{max}$ as observing an object and quit algorithm, else go to step 1.

The distributed algorithm described above leads to the same decision rule as the centralized BH procedure. However when there is a communication constraint of $\alpha$ bits for the SNET, we only need to put a cap on the $count$ variable and perform the distributed BH algorithm while $count_t \leq \alpha$.

The aforementioned distributed algorithm is based on linear increase in threshold at each step. This leads to FDR control at level $\gamma m_0/m$. This leads to inherent conservatism in cases where the number of objects is a finite non-zero fraction. We have analyzed an alternative strategy in [10]. Therein, at each count update step an estimate of actual number of targets, $\hat{m}_1$, is computed based on the number of sensors declared as significant. The threshold $l(i_t)$ is then adjusted based on the estimated target density. Our simulation results indicated that this strategy leads to a much better detection power, as seen in Figure 11.

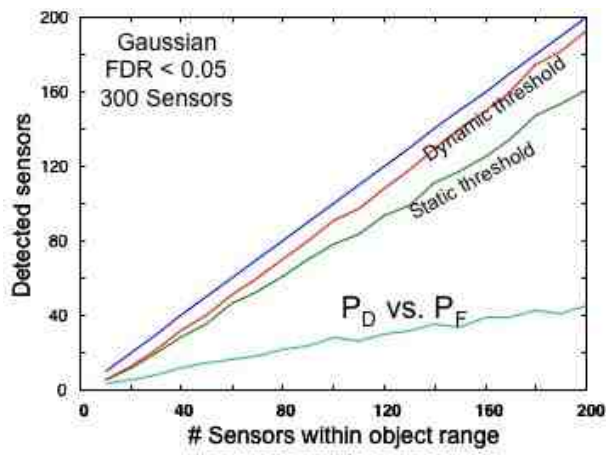

Figure 11: Constraint update through learning: The estimate of $m_0$ allows for update of threshold $\gamma$ for better detection rates.

**Scaling Property:** We now establish the simple but very important scaling property of the distributed DTBH procedure. It is due to this property that we can limit the communication budget with an upper bound that depends on the number of sensors that have an object within their sensing range, and not the total number of sensors in the SNET.

**Theorem 3.10** *Let $m_1 = m - m_0$ be the number of sensors with positive hypothesis. The expected ratio of $m_1$ to the number of sensors that are declared to be significant (R) is lower bounded by $1 - \gamma$; i.e. $E\{m_1/R\} \geq 1 - \gamma$.*

Proof: We know that the BH procedure guarantees $E\{V/R\} = E\{V/(V+S)\} \leq \gamma$. Evidently $S \leq m_1$, meaning the number of correct detections cannot exceed the number of sensors with positive hypotheses. Then:

$$E\{V/(V+S)\} = 1 - E\{S/(V+S)\} \Rightarrow E\{S/(V+S)\} \geq 1 - \gamma$$

$$1 - \gamma \leq E\{S/(V+S)\} \leq E\{m_1/(V+S)\} = E\{m_1/R\}$$

which concludes the proof of this theorem. ∎

### 3.4.1 Distributed DTBH achieves Performance of Last Crossing Procedure

Note that the distributed algorithm is a first-crossing procedure. Once the threshold line crosses below the ordered $p$ values, the algorithm terminates. The issue is that to maintain FDR control it is only required to terminate at the last crossing. Therefore, there could be degradation in performance if one terminates at first crossing. When there is an asymptotically large number of observations, the ordered $p$ values form a convex function. In that case, the first-crossing and the last-crossing procedures are the same, and they terminate at the same point. However, when there is not enough samples, the ordered $p$ values do not form a convex function, and therefore the first-crossing and the last-crossing procedures have different termination points. Below we show that the above procedure achieves the last-crossing performance with high probability.

Let the largest $p$ value that is below its corresponding threshold be $p'$. For this part, let $\theta_0 = m_0/m$, $\theta_1 = m_1/m$, be the ratio of observations from each hypothesis. We use the definition of $l(i) = i\gamma/m, \ i = 1\ldots m$ as the threshold line for the centralized BH procedure. We assume that the PDF of $P_1$ is monotonically decreasing due to the domain transformation.

**Lemma 3.11** *Let $p_k$ be the $k^{th}$ smallest p value. If $E(p_{\lceil \frac{k}{1-\epsilon} \rceil}) \leq l_k$, then $Pr\{p_k > l_k\}$ decays exponentially fast with $k$.*

Proof: Let $N_k = \#\{j : p_j \leq l_k\} = \sum_{j=1}^{m} I_{\{p_j \leq l_k\}}$. By the switching lemma (see for example [1]) the following relationship holds for any $k$: $\{E(p_{\lceil \frac{k}{1-\epsilon} \rceil}) \leq l_k\} \Leftrightarrow \{E(N_k) \geq \lceil \frac{k}{1-\epsilon} \rceil\}$. Therefore, $E(p_{\lceil \frac{k}{1-\epsilon} \rceil}) \leq l_k \Rightarrow E(N_k) \geq$

$\frac{k}{1-\epsilon}$ and $k \leq E(N_k)(1-\epsilon)$.

$$\begin{aligned}
\Pr\{p_k > l_k\} = \Pr\{N_k < k\} &\leq \Pr\{N_k < E(N_k)(1-\epsilon)\} \\
&\leq \exp\{-\frac{\epsilon^2 E(N_k)}{2}\} \quad (2) \\
&\leq \exp\{-\frac{\epsilon^2 k}{2(1-\epsilon)}\} \quad (3)
\end{aligned}$$

Inequality 2 follows from the Chernoff bound, and inequality 3 follows from the application of switching relation along with the assumption of the theorem. ∎

The implication of this lemma is that after a certain number of $p$ values, say $k$, are tested against their corresponding thresholds, one can decide whether or not to continue the distributed algorithm with an exponentially small probability of error.

This result further suggests presetting $k$ tests at the beginning of the algorithm, which must be performed regardless of the outcome. Note, however, that $k$ can be fixed a priori and does not depend on the size of the SNET. We next show that such a modification does not affect important properties of our distributed algorithm.

**Theorem 3.12** *Consider the distributed detection algorithm with $k$ preset tests. For that distributed implementation:*
**a)** $FDR \leq \gamma$ *and*
**b)** *The expected number of bits required to detect the objects is upper bounded by* $\max\{k, E(\frac{m_1}{1-\gamma})\}$

Proof: **a)** We show this part by showing that the distributed algorithm is in fact equivalent to the centralized algorithm, and that presetting $k$ tests affects only the communication cost. If there exists a $p_i \leq i\gamma/m$, $i \geq k$, then the effect of $k$ preset tests is washed out. This is because the centralized algorithm would also declare all $p$ values less than $p_i$ significant. Therefore $FDR \leq \gamma$ in this case. If there is no such $p_i$, $i \geq k$, then the algorithm chooses the largest $p$ value $p_j \leq j\gamma/m$, $j < k$, and declares all smaller $p$ values significant. Therefore the distributed algorithm is equivalent to the centralized FDR procedure, hence $FDR \leq \gamma$.
**b)** Without the $k$ preset tests the upper bound was given by Theorem 3.10, and the result is immediate from there. ∎

## 4 Robustness of DTBH and Non-Ideal Sensing Model

The main results of this section are directed toward control of FDR via DTBH procedure when the distribution of observations under null hypotheses are not known exactly. The reasoning for developing robustness in this perspective is as follows: in the non-ideal sensing model the sensors that are outside the effective region receive a small residual signal from the objects. Since the received signal is not known, the exact distribution of observations under null hypotheses are not available to calculate the $p$ values. Furthermore, in multidimensional settings it may not be possible to obtain the level sets exactly, and this introduces an uncertainty in the distributions that are used to perform the domain transformation. This leads to a deviation of $P_0$ distribution from $U(0,1)$. Our goal here is to quantify how FDR is affected when DTBH is used with non-ideal sensing model.

For this section we again assume that an appropriate definition of $p$ values has been chosen, and that we are working in the $p$ space as opposed to the original observation space. Further, we assume that a domain transformation $\mathcal{T}$ has been chosen that preserves uniform distribution. Then we establish three main results. First, we show that FDR scales gracefully when $P_0$ distribution deviates from $U(0,1)$ by $\epsilon$ under a suitable metric. Next, we show that the domain transformation preserves the distance $\epsilon$ between the distribution of $\hat{P}_0$ and $U(0,1)$. These results allow us to identify one topology in which our developments can address the distributed detection problem with non-ideal sensing model. We then show how the non-ideal sensing model can be addressed with the proposed method.

For continuous families $\mathcal{F}_0$ such that $\mathcal{F}_0 = \{F_0 : |F_0(x) - x| \leq \epsilon x\}$, we have an immediate non-asymptotic robustness result, which states that the FDR scales gracefully when BH procedure is used for detection.

**Lemma 4.1** *Let $P_0$ have continuous distribution $F_0(x)$. If $|F_0(x) - x| \leq \epsilon x$, the BH procedure bounds the false discovery rate by $\gamma(1+\epsilon)$, i.e. $FDR \leq \gamma(1+\epsilon)$.*

Proof: Define $\gamma_i = i\gamma/m$. Let $P_{0i}$ be the $m_0$ $p$ values. Denote with $C_i(k)$ the event that if $p_i$ is declared $H_1$, exactly $k - 1$ other $p$ values are declared $H_1$. Then;

$$\begin{aligned}
E(V/R) &= \sum_{i=1:m_0} \sum_{k=1:m} \frac{1}{k}\Pr\{P_{0i} \leq \gamma_k, C_i(k)\} = \sum_{i=1:m_0} \sum_{k=1:m} \frac{1}{k}\Pr\{P_{0i} \leq \gamma_k\}\Pr\{C_i(k)\} \\
&\leq \sum_{i=1:m_0} \sum_{k=1:m} \frac{1}{k}\frac{\gamma k}{m}(1+\epsilon)\Pr\{C_i(k)\} = (1+\epsilon) \sum_{i=1:m_0} \frac{\gamma}{m} \sum_{k=1:m} \Pr\{C_i(k)\} \\
&= (1+\epsilon) \sum_{i=1:m_0} \frac{\gamma}{m} = (1+\epsilon)\frac{\gamma m_0}{m} \leq (1+\epsilon)\gamma
\end{aligned}$$

The second equality follows because $P_{0i}$ is independent of all other $p$ values. ∎

In order to extend Lemma 4.1 to the DTBH procedure, we need to show that the distance $\epsilon$ is preserved when we apply the domain transformation. The following lemma states this result.

**Lemma 4.2** *Let $P_0$ have continuous distribution $F_0(x)$ and $\hat{P}_0 = \mathcal{T}[P_0]$ have continuous distribution $\hat{F}_0(x)$. If $|F_0(x) - x| \leq \epsilon x$, then $|\hat{F}_0(x) - x| \leq \epsilon x$.*

Proof: The result is an immediate consequence of the fact that $\mathcal{T}$ is a many to one mapping only over countable sets. ∎

Combining the results of Lemma 4.1 and Lemma 4.2 we state the main robustness property of the DTBH procedure without the obvious proof.

**Theorem 4.3** *For continuous families $\mathcal{F}_0$ such that $\mathcal{F}_0 = \{F_0 : |F_0(x) - x| \leq \epsilon x\}$, the DTBH procedure bounds the false discovery rate by $\gamma(1 + \epsilon)$, i.e. $FDR \leq \gamma(1 + \epsilon)$.*

Before going further we note that Lemma 4.2 can be extended to families of size $\epsilon$ in Kolmogorov or Prokhorov metrics, and these extensions allow us to consider singular distributions as well as continuous ones. The distributed detection algorithm can be modified to accommodate for these more general families of distributions, which leads to a variant of Theorem 4.3. We present the proof of the extension of Lemma 4.2 to Kolmogorov metric in the appendix and omit the modification of the distributed detection algorithm as well as development of the variant of Theorem 4.3 for brevity.

**Theorem 4.4** *Let $\mu$ be the measure associated with $F(x) = x$ and $\mu_0$ be the measure associated with $F_0$. Define $\hat{F}$ and $\hat{F}_0$ to be the respective distributions after the transformation. If $d_{tv}\{\mu, \mu_0\} \leq \epsilon$ then $\sup_x \mid \hat{F}(x) - \hat{F}_0(x) \mid \leq \epsilon$ where for any measurable space $\Omega$, $d_{tv}(\mu, \nu) = \sup_{A \subset \Omega} \mid \mu(A) - \nu(A) \mid$.*

**Non-Ideal Sensing Model:** The robustness result stated in Theorem 4.3 presents us with an immediate modification to the DTBH algorithm in order to control the false discovery rate. It suggests that if we wish to control FDR at level $\gamma$, we only need to input the threshold $\gamma' = \gamma/(1 + \epsilon)$. Then the distributed DTBH algorithm presented in Section 3.4 can address the problem with Non-Ideal Sensing Model, however with a performance loss. Here we present only the form of this loss, as it depends on distribution specific values.

Let $\hat{P}_1$ have concave distribution $\hat{F}_1(x)$. In [11] it has been shown that asymptotically the decision point of the BH procedure is $c$, where $c$ is the solution to

$$\hat{F}_1(x) = \frac{1/\gamma - m_0/m}{m_1/m} x \qquad (4)$$

Asymptotically, this yields $E(T)_\gamma = (1 - \hat{F}_1(c))m_1$, which would have been the solution in the Ideal Sensing Model. Since there is uncertainty in the family $\mathcal{F}_0 = \{F_0 : |F_0(x) - x| \leq \epsilon x\}$, we use the new threshold in the DTBH procedure: $\gamma' = \gamma/(1 + \epsilon)$. This threshold will yield a new point $c'$ such that $c'$ is the solution to

$$\hat{F}_1(x) = \frac{1/\gamma' - m_0/m}{m_1/m} x = \frac{(1+\epsilon)/\gamma - m_0/m}{m_1/m} x \qquad (5)$$

We note that $c' < c$ since the slope of the right hand side of equation 5 is larger than that of equation 4. The new threshold yields a new miss rate, i.e. $E(T)_{\gamma'} = (1 - \hat{F}_1(c'))m_1$. The performance loss is then directly related to $c$ and $c'$ via the function $\hat{F}_1(x)$: $E(T)_{\gamma'} - E(T)_\gamma = (\hat{F}_1(c) - \hat{F}_1(c'))m_1$.

## 5  Simulation Results

Below we present a detection simulation in which we use BH and DTBH procedures. The sensor field for the simulations is a grid of size 100x100, where each pixel is assumed to have a sensor, and the sensors observe the signal within their pixel. The null hypotheses for a sensor is that it is outside the effective region of all objects, and the alternative is that it is inside the effective region of an object. Then, with $n_s$ and $\nu_s$ being noise at sensor $s$ for null and alternative hypotheses respectively, the observation model at sensor $s$ for the non-ideal case is as follows:

$$H_0 \; : \; X_s = \xi_s + n_s$$
$$H_1 \; : \; X_s = \theta_s + \nu_s$$

In the ideal sensing model, $\xi_s = 0$ and $\theta_s = \theta$. In the non-ideal sensing model, $\xi_s \in [0, 0.1]$, $\theta_s \in [\theta - 0.1, \theta]$. Here we have absorbed the perturbation terms to $\xi_s$, and $\theta_s$. We chose a demonstrative distribution for $\nu_s$ and a demonstrative value for $\theta$. We note that similar results are obtained when these values are varied. In cases where $\theta$ is smaller, the detection rate of DTBH method remains the same, whereas the detection rate of BH procedure degrades significantly.

The results demonstrate the robustness of DTBH procedure to such non-ideal sensing scenarios. For the simulation, the FDR threshold was set to $\gamma = .15$, $\theta = 2.8$, the effective radius of the object $r_{eff} = 2.5$ pixels. $n_s \sim N(0, 1)$ and $\nu_s \sim N(0, 0.05)$. There were 10 objects on the field. The communication constraint $\alpha$ was varied and the results are presented for illustrative cases in Figure 12 and Figure 13 for the ideal and non-ideal sensing models respectively.

In the ideal sensing model, for $\alpha \leq 150$ implementation of the BH procedure was unable to detect the sensors with $H_1$ hypotheses, whereas the DTBH procedure was able to do so. As the communication constraint was loosened, the performance of DTBH procedure increased accordingly, yet keeping the false alarms at low levels. Although the BH procedure also detected some sensors with $H_1$ hypotheses, observe that the BH procedure suffers from more false alarms.

In the non-ideal sensing model, note that the BH procedure fails to detect with any amount of communication budget. This is because the ordered $p$ values are always above their corresponding thresholds. However, although the exact distribution is not known under positive hypotheses, the domain transformation is performed successfully, and this allows for successful detection of significant sensors.

# 6  Appendix

**Proof of Theorem 1** First note that from Lagrangian duality it follows that, $\gamma_w \geq \gamma_b$. Consequently, we are left to establish a bound for the Bayesian problem. Now we observe that the error event,

$$E = \{u(Y^N) \neq \{H^s : s \in \mathcal{S}\}\} = \{V \geq 1\} \cup \{T \geq 1\}$$

Therefore, from Fano's inequality it follows that for any strategy $\phi$:

$$\Pr(V \geq 1) + \Pr(T \geq 1) \geq \Pr(E) \geq \frac{1}{N}\Phi\{H^s : s \in \mathcal{S}\} \mid Y^N) - \frac{1}{N} = \Phi(H^s \mid Y_s) - \frac{1}{N}$$

The last equality follows from the independence assumptions. ∎

**Technical Details of Domain Transformation:**

**Proposition 6.1** $\hat{F}_1(\alpha_\mu(y)) = \hat{\mu}(0, \alpha_\mu(y)) = \beta_\mu(y)$ *is a distribution function.*

Proof: To show that $\hat{F}_1(\cdot)$ is a distribution function, we need to show that **(1)** $\hat{F}_1(\cdot)$ is monotone increasing, **(2)** $\hat{F}_1(\cdot)$ is right continuous, **(3)** $\lim_{x\to-\infty} \hat{F}_1(x) = 0$ and $\lim_{x\to+\infty} \hat{F}_1(x) = 1$.

**(1)** $\hat{F}_1(\alpha_\mu(y))$ is monotone increasing in $\alpha_\mu(y)$. This is because $\alpha_\mu(y)$ increases as $y$ decreases, and as $y$ decreases $\beta_\mu(y) = E_\mu[I_{\{f_1(x) \geq y\}}(x)]$ increases.

**(2)** Next, $\hat{F}_1(\alpha_\mu(y))$ is a right-continuous function. To show this, consider some $\alpha'$, and appropriate $y'$ and $\beta'$. $\alpha_\mu(y) \downarrow \alpha'$ as $y \uparrow y'$. But, $y \uparrow y' \Rightarrow \beta_\mu(y) \downarrow \beta'$, and since $E_\mu[I_{\{f_1(x) \geq y\}}(x)]$ is right continuous so is $\hat{F}_1(\alpha_\mu(y)) = \beta_\mu(y)$.

**(3)** Finally, we show that $\lim_{\alpha_\mu(y) \to 0} \hat{F}_1(\alpha_\mu(y)) = 0$ and $\lim_{\alpha_\mu(y) \to 1} \hat{F}_1(\alpha_\mu(y)) = 1$. The reason that we only consider 0 and 1 as the limit points is because $\alpha_\mu(y) \in [0, 1]$ by definition.

Note that since $f_1$ is the PDF of a continuous random variable, there exists a $y'$ such that $\{x : f_1(x) \geq y'\}$ is empty. Therefore, as $y \uparrow y'$, $\alpha_\mu(y) \downarrow \alpha_\mu(y') = 0$, and $\beta_\mu(y) \downarrow \beta_\mu(y') = 0$. This shows the first part. Next, consider the case when $y \downarrow 0$. In this case $\alpha_\mu(y) \uparrow \alpha_\mu(0) = 1$ and $\beta_\mu(y) \uparrow \beta_\mu(0) = 1$. This establishes that $\hat{F}_1(\alpha_\mu(y))$ is a distribution function. ∎

We now show that $\hat{\mu}$ is absolutely continuous with respect to Lebesgue measure, thus admits a PDF.

**Proposition 6.2** $\hat{\mu}$ *is absolutely continuous with respect to Lebesgue measure.*

Proof: Over any zero measure set $A$ with respect to Lebesgue measure, $\alpha_\mu(y) = 0$ and $\beta_\mu(y) = 0$ and therefore $\hat{\mu}(A) = 0$. ∎

**Proof of Proposition 3.5** Note that $\beta_\mu(y)$ is concave as a function of $\alpha_\mu(y)$. To show this consider $y_1 > y_2 > \ldots > y_n$ and a sequence of sets $A_1 \subset A_2 \subset \ldots \subset A_n$ such that $A_i = \{x : f_1(x) \geq y_i\}$. Note that $\mathcal{L}(A_1) \leq \mathcal{L}(A_2) \leq \ldots \leq \mathcal{L}(A_n)$ where $\mathcal{L}$ denotes the Lebesgue measure. Since $\sup_{x \in A_{i+1} - A_i} f_1(x) \leq \inf_{x \in A_i} f_1(x)$, $d\beta_\mu/d\alpha_\mu$ is monotonically decreasing in these sets. Therefore, the new measure $\hat{F}_1(\alpha_\mu) = \hat{\mu}(0, \alpha_\mu)$ is concave, and the proposition follows. ∎

**Proof of Theorem 3.7** Let $f_0(\cdot)$ and $f_1(\cdot)$ be PDFs of observations under $H_0$ and $H_1$ respectively. Let $\mathcal{P} = \{P_1, P_2, \ldots, P_m\}$ be given, where $P_i$ are independent random variables having PDFs $f_0$ or $f_1$ with unknown prior probabilities $\Pr\{H^i = H_0\}$ and $\Pr\{H^i = H_1\}$ respectively. Now consider the partitioning problem as described in Definition 3.6.

Let $S$ be a decision rule that chooses $\mathcal{P}_S \subset \mathcal{P}$ and labels $H_1$. Define $P^* = \max_i \{P_i \in \mathcal{P}_S\}$ and $P_* = \min_i \{P_i \in \mathcal{P}_S^c\}$ where $\mathcal{P}_S^c = \mathcal{P} - \mathcal{P}_S$. Now define a new decision rule $S'$ as follows: If $\mathcal{P}_s \neq \phi$ and $P_* \leq p < P^*$ for some $p$, then $S'$ chooses $\mathcal{P}_{S'} = (\mathcal{P}_S - \{P^*\}) \cup \{P_*\}$ and labels $H_1$. In all other cases $S'$ chooses $\mathcal{P}_{S'} = \mathcal{P}_s$. With this setup, the following lemma establishes that strategy $S'$ suffers a smaller FDR than does strategy $S$.

**Lemma 6.3** *Let relevant quantities be defined as above. Assume that $Pr\{H^i = H_0\} = Pr\{H^j = H_0\}$, $\forall i, j : 1, \ldots, m$. If $f_0(\cdot) = U(0,1)$ and $f_1(\cdot)$ is a monotonically decreasing PDF, then the false discovery rate of the strategy $S'$ is less than or equal to that of the strategy $S$, i.e., $FDR_S \geq FDR_{S'}$.*

Proof: Let $m = m_0 + m_1$ and let $\omega \in (0,1)^m$. Let $B = \{\omega : P_* \leq p, P^* > p\}$ have a nonzero measure. Outside $B$, $S$ itself is a thresholding strategy and $S = S'$. Therefore we only need to consider the set $B$. Let $i^*(\omega)$ and $i_*(\omega)$ be the indices of $P^*$ and $P_*$ in the set $\mathcal{P}$. Now, consider a $B' \subset B$ in which $i^*(\omega) = i^*$ and $i_*(\omega) = i_*$ are fixed.

Define $D_S(i) = I(P_i \in \mathcal{P}_S)$, $D_{S'}(i) = I(P_i \in \mathcal{P}_{S'})$, and $A(i) = I(H^i = H_0)$. Then,

$$\begin{aligned} FDR_S - FDR_{S'} &= E(\frac{V_S}{R_S}|B') - E(\frac{V_{S'}}{R_{S'}}|B') \\ &= E(\frac{\sum_{i=1:m} D_S(i)A(i)}{\sum_{i=1:m} D_S(i)}|B') - E(\frac{\sum_{i=1:m} D_{S'}(i)A(i)}{\sum_{i=1:m} D_{S'}(i)}|B') \\ &= E(\frac{\sum_{i=1:m} D_S(i)A(i)}{\sum_{i=1:m} D_S(i)} - \frac{\sum_{i=1:m} D_{S'}(i)A(i)}{\sum_{i=1:m} D_{S'}(i)}|B') \end{aligned}$$

Now, note that the cardinality of $\mathcal{P}_S$ and $\mathcal{P}_{S'}$ are the same, and $D_S(i) \neq D_{S'}(i)$ only for $i^*$ and $i_*$. Therefore we can rewrite the above difference as follows:

$$\begin{aligned}
FDR_S - FDR_{S'} &= E(\frac{\sum_{i=1:m, i\neq\{i^*,i_*\}} D_S(i)A(i) + D_S(i^*)A(i^*) + D_S(i_*)A(i_*)}{\sum_{i=1:m} D_S(i)}|B') \\
&- E(\frac{\sum_{i=1:m, i\neq\{i^*,i_*\}} D_{S'}(i)A(i) - D_{S'}(i^*)A(i^*) - D_{S'}(i_*)A(i_*)}{\sum_{i=1:m} D_S(i)}|B') \\
&= E(\frac{D_S(i^*)A(i^*) - D_{S'}(i_*)A(i_*)}{\sum_{i=1:m} D_S(i)}|B') = E(\frac{A(i^*) - A(i_*)}{\sum_{i=1:m} D_S(i)}|B')
\end{aligned}$$

$$\begin{aligned}
FDR_S - FDR_{S'} &\geq \frac{E(A(i^*) - A(i_*)|B')}{m} = \frac{E(I(H^{i^*} = H_0) - I(H^{i_*} = H_0)|B')}{m} \\
&= \frac{\Pr\{H^{i^*} = H_0|B'\} - \Pr\{H^{i_*} = H_0|B'\}}{m} \\
&= \frac{1}{m}[\frac{\Pr\{H^{i^*} = H_0, B'\}}{\Pr\{B'\}} - \frac{\Pr\{H^{i_*} = H_0, B'\}}{\Pr\{B'\}}] \\
&= \frac{1}{m}[\frac{\Pr\{B'|H^{i^*} = H_0\}\Pr\{H^{i^*} = H_0\}}{\Pr\{B'\}} - \frac{\Pr\{B'|H^{i_*} = H_0\}\Pr\{H^{i_*} = H_0\}}{\Pr\{B'\}}] \\
&= \frac{1}{m}[\frac{\Pr\{P^* > p|H^{i^*} = H_0\}\Pr\{P_* \leq p\}\Pr\{H^{i^*} = H_0\}}{\Pr\{P^* > p\}\Pr\{P_* \leq p\}} \\
&- \frac{\Pr\{P_* \leq p|H^{i_*} = H_0\}\Pr\{P^* > p\}\Pr\{H^{i_*} = H_0\}}{\Pr\{P^* > p\}\Pr\{P_* \leq p\}}]
\end{aligned}$$

Here observe that $\Pr\{H^{i^*} = H_0\} = \Pr\{H^{i_*} = H_0\}$ by the hypothesis of the theorem, and $\Pr\{P_* \leq p\} = 1 - \Pr\{P^* > p\}$ due to independence assumption. Therefore,

$$\begin{aligned}
FDR_S - FDR_{S'} &\geq \frac{\Pr\{H^{i^*} = H_0\}}{m}[\frac{(1-p)(1 - \Pr\{P^* > p\})}{\Pr\{P^* > p\}\Pr\{P_* \leq p\}} - \frac{p\Pr\{P^* > p\}}{\Pr\{P^* > p\}\Pr\{P_* \leq p\}}] \\
&= \frac{\Pr\{H^{i^*} = H_0\}}{m\Pr\{P^* > p\}\Pr\{P_* \leq p\}}[1 - \Pr\{P^* > p\} - p + p\Pr\{P^* > p\} - p\Pr\{P^* > p\}] \\
&= \frac{\Pr\{H^{i^*} = H_0\}}{m\Pr\{P^* > p\}\Pr\{P_* \leq p\}}[F_{P^*}(p) - p] \geq 0
\end{aligned}$$

The last inequality follows from the fact that $f_0 = U(0,1)$ and $f_1$ is monotonically decreasing. ∎

Now, to prove the result of Theorem 3.7 it suffices to iterate Lemma 6.3 whenever the set $S'$ is not a threshold set, i.e. the result of a threshold strategy. Specifically, if $S'$ is not a threshold set, redefine $S = S'$, generate a new $S'$, and repeat this procedure until $S'$ is a threshold set. ∎

**Proof of Lemma 3.8**

$$f_{X_{(i)}}(t) = \frac{n!}{(i-1)!(n-i)!}(F_X(t))^{i-1}(1 - F_X(t))^{n-i}f_X(t)$$

$$F_{X_{(i)}}(t) = \frac{n!}{(i-1)!(n-i)!} \int_0^t (F_X(x))^{i-1}(1-F_X(x))^{n-i} f_X(x) dx$$

$$= \frac{n!}{(i-1)!(n-i)!} \int_0^{F_X(t)} u^{i-1}(1-u)^{n-i} du$$

By the same approach it is easy to see that

$$F_{Y_{(i)}}(t) = \frac{n!}{(i-1)!(n-i)!} \int_0^t (F_Y(y))^{i-1}(1-F_Y(y))^{n-i} f_Y(y) dy$$

$$= \frac{n!}{(i-1)!(n-i)!} \int_0^{F_Y(t)} u^{i-1}(1-u)^{n-i} du$$

By hypothesis of the lemma, $F_X(t) \geq F_Y(t) \ \forall t \in (0,1)$, and since $0 \leq u \leq 1$, it follows that $F_{X_{(i)}}(t) \geq F_{Y_{(i)}}(t)$ $\forall t \in (0,1)$, which concludes the proof of the lemma. ∎

**Proof of Theorem 4.4**

Proof: Let $A_x \subset (0,1)$ be the set that gets mapped to the set $(0,x)$ by the transformation. Since $d_{tv}\{\mu, \mu_0\} \leq \epsilon$, by definition of total variation distance we have $|\mu(A_x) - \mu_0(A_x)| \leq \epsilon$. Noting that $\mu(A_x) = \hat{\mu}(0,x)$ and $\mu_0(A_x) = \hat{\mu}_0(0,x)$ we have $|\hat{\mu}(0,x) - \hat{\mu}_0(0,x)| \leq \epsilon$ and the result follows. ∎

# References

[1] F. Abramovich, Y. Benjamini, D. Donoho, and I. Johnstone. Adapting to unknown sparsity by controlling the false discovery rate. *Technical Report, Statistics Department, Stanford University*, 2000.

[2] M. Alanyali, S. Venkatesh, O. Savas, and S. Aeron. Distributed bayesian hypothesis testing in sensor networks. In *American Control Conference*, Boston, MA, July 2004.

[3] S. Appadwedula, V. V. Veeravalli, and D. L. Jones. Robust and locally-optimum decentralized detection with censoring sensors. In *5th Int. Conf. on Information Fusion*, 2002.

[4] Y. Benjamini and Y. Hochberg. Controlling the false discovery rate: A practical and powerful approach to multiple testing. *Journal of the Royal Statistical Society, Series B*, 57, 1995.

[5] Y. Benjamini and D. Yekuteli. The control of the false discovery rate in multiple testing under dependency. *The Annals of Statistics*, 29, 2001.

[6] J.F. Chamberland and V. Veeravalli. Asymptotic results for decentralized detection in power constrained wireless sensor networks. *IEEE Special Issue on Wireless Sensor Networks*, pages 1007–1015, August 2004.

[7] Z. Chi. False discovery rate control with multivariate p-values. Technical Report TR 06-10, University of Connecticut Department of Statistics.

[8] E. Ermis and V. Saligrama. Adaptive sampling strategies for detection of localized phenomena, information processing in sensor networks. In *4th International Workshop on Information Processing in Sensor Networks*, Los Angeles, CA, 2005.

[9] E. Ermis and V. Saligrama. Detection and localization in sensor networks using fdr. In *Conference on Information Sciences and Systems*, Princeton, NJ, 2006.

[10] E. Ermis and V. Saligrama. Dynamic thresholding for distributed multiple hypotheses testing. In *IEEE Statistical Signal Processing Workshop*, Madison, WI, 2007.

[11] C. Genovese and L Wasserman. Operating characteristics and extensions of the false discovery rate procedure. *Journal of the Royal Statistical Society, Series B*, 64:479–498, 2002.

[12] T. Kasetkasem and P. K. Varshney. Communication structure planning for multisensor detection systems. *IEE Proc. Radar, Sonar Navig.*, 48:2–8, 2001.

[13] E. L. Lehmann. *Testing Statistical Hypothesis*. John Wiley and Sons, 1986.

[14] M. Longo, T. D. Lookabaugh, and Robert M. Gray. Quantization for decentralized hypothesis testing under communication constraints. *IEEE Trans. on Inform. Theory*, 36:241–255, 1990.

[15] M. Alanyali O. Savas and V. Saligrama. Efficient in-network processing through information coalescence. In *Information Processing in Sensor Networks*, submitted, 2006.

[16] N. Patwari and A. Hero. Reducing transmissions from wireless sensors in distributed detection networks using hierarchical censoring. In *ICASSP*. IEEE, 2003.

[17] C. Rago, P. Willett, and Y. Bar-Shalom. Censoring sensors: a low-communication-rate scheme for distributed detection. *IEEE Trans. Aerosp. Electron. Syst.*, 32:554–568, 1996.

[18] S. Ross. *Introduction to Stochastic Dynamic Programming*. Academic Press, 1983.

[19] J. D. Storey. A direct approach to false discovery rates. *Journal of the Royal Statistical Society, Series B*, 64:499–517, 2002.

[20] J. D. Storey. The positive false discovery rate: a bayesian interpretation and the q-value. *Annals of Statistics*, 31:2013–2035, 2003.

[21] P.F. Swaszek and P. Willett. Parley as an approach to distributed detection. *IEEE Transactions on Aerospace and Electronic Systems*, pages 447–457, January 1995.

[22] J. N. Tsitsiklis. Decentralized detection. *in Advances in Statistical Signal Processing, H. V. Poor and J. B. Thomas Eds*, 2, 1993.

[23] V. Saligrama, M. Alanyali, O. Savas. Distributed detection in sensor networks with packet losses and finite capacity links. *IEEE Transactions on Signal Processing*, November 2006.

[24] P. K. Varshney. *Distributed Detection and Data Fusion*. Springer, 1997.

[25] William C. Karl Venkatesh Saligrama, Yonggang Shi. Performance guarantees in sensor networks. *IEEE Int. Conf. on Acoust., Speech, and Sig. Process.*, 2004.

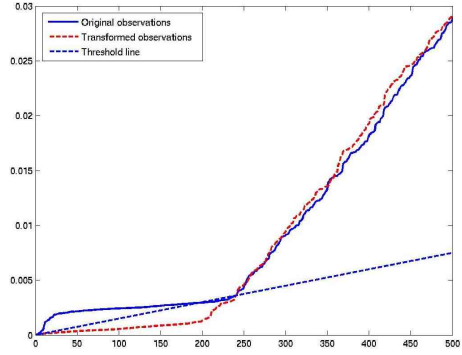

(a) Ordered $p$ values

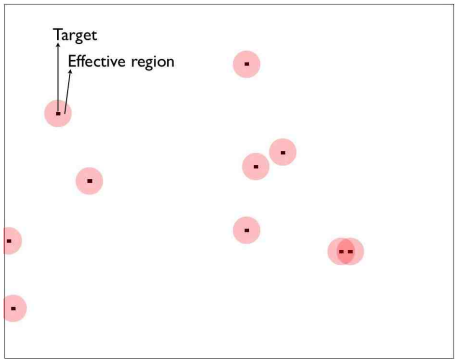

(b) Object Locations

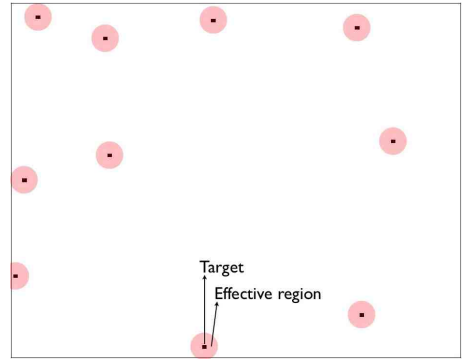

(c) Object Locations

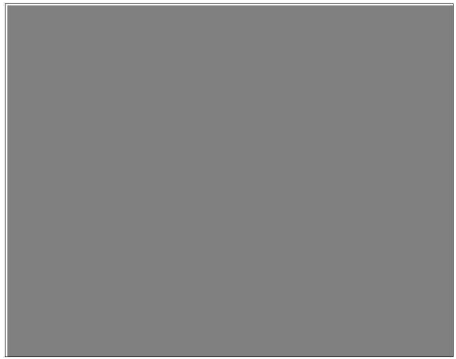

(d) Detection with BH

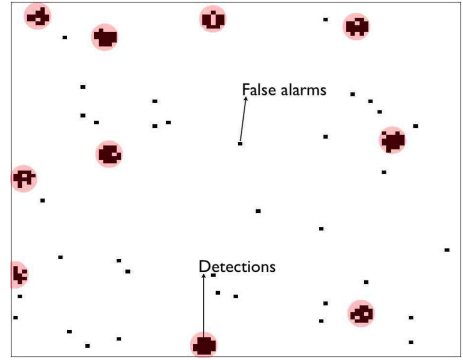

(e) Detection with BH

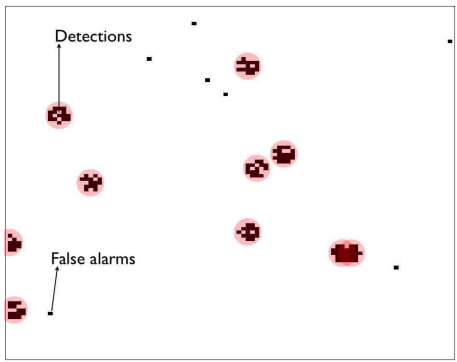

(f) Detection with DTBH

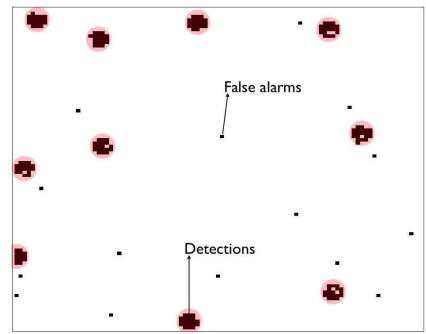

(g) Detection with DTBH

Figure 12: Detection performance of distributed implementations under ideal sensing model for $\alpha = 150$ bits (b,d,f), $\alpha = 200$ bits (c,e,g). (A purely gray plot indicates that no detection was made.)

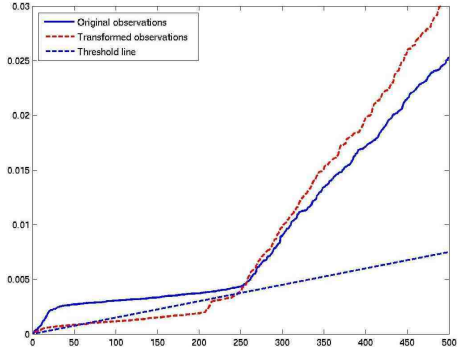

(a) Ordered $p$ values

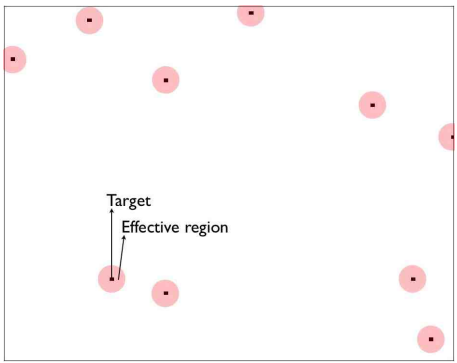

(b) Object Locations

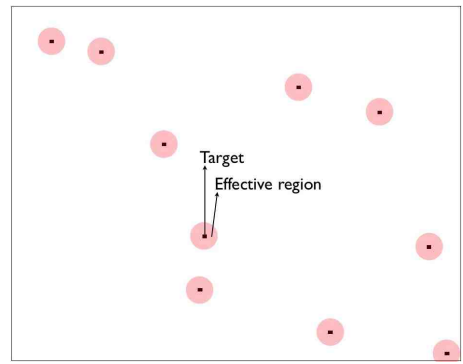

(c) Object Locations

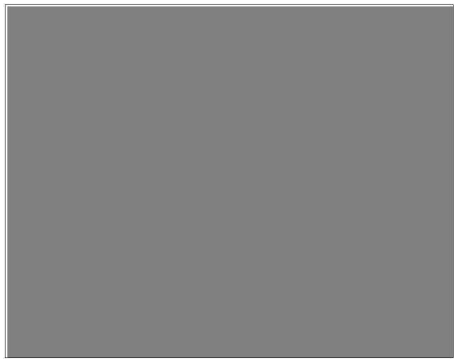

(d) Detection with BH

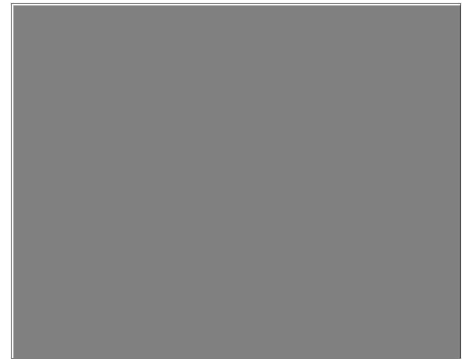

(e) Detection with BH

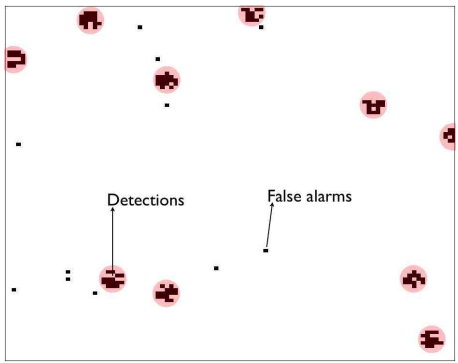

(f) Detection with DTBH

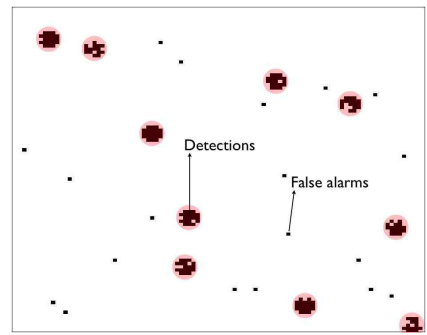

(g) Detection with DTBH

Figure 13: Detection performance of distributed implementations under non ideal sensing model for $\alpha = 150$ bits (b,d,f), $\alpha = 200$ bits (c,e,g). (A purely gray plot indicates that no detection was made.)